# Application of Thomas-Fermi model to a negative hydrogen ion in a strong electric field

Yuri Kornyushin
*Maître Jean Brunschvig Research Unit, Chalet Shalva*
*Randogne, 3975-CH*

Thomas-Fermi model is applied to describe some basic properties of a negative hydrogen ion in a strong electric field. The equilibrium ionic size, energy and polarizability of the ion are calculated. Collective modes of the dipole oscillations are regarded. A barrier, due to which the ion is in a stable state, is studied. The barrier vanishes at some large value of the electric field, which is defined as a critical value. The dependence of the critical field on frequency is studied. At high frequencies a 'stripping' mechanism for instability arises. At the resonant frequency very low amplitude of the electric field causes the 'stripping' instability.

## 1. Introduction

There is a strong attractive center in atom, the nucleus, which acts on the electrons in an atom and to a large extent determines all the atomic properties. The electrons occupy given energy levels, forming shells and subshells. All this is understandable without accounting for the interelectron interaction. However, although not being able to eliminate the shell structure, the interelectron interaction modifies it considerably, bringing large corrections even to the total binding energy in the medium-range and many-electron heavy atoms. These corrections are of the same order of magnitude as the Coulombic interaction of the nucleus with the electrons, but are, of course, smaller.

Two approaches were developed for atoms. One of them is based on the assumption that the electrons in the atom are predominantly independent, and move in a so-called self-consistent field, which is formed by the potential of the nucleus and the electronic potentials, created by averaging of the interelectron interaction over the space positions of all the atomic electrons acting upon the one under consideration. This approach was suggested by Hartree [1] and considerably improved by Fock [2], who included exchange between the indistinguishable atomic electrons. Generally speaking, Hartree-Fock method presents the best possible approach, in which electrons are considered as moving inside the atom independently. In the framework of this approach and its natural generalization to excited states, a number of essential features of the atomic structure, including the binding energies, were described reasonably well. In order to do this, extensive and rather complicated numerical calculations had to be performed.

However, it appears that Hartree-Fock approach fails to tell 'the whole truth' about atoms, and there are a number of characteristics, such as the excitation spectra, photoionization cross-sections, etc, for which Hartree-Fock method does not yield good results and sometimes even gives qualitatively incorrect ones. It turns out that the direct interelectronic interaction, which leads to the distortions of independent motion of electrons inside the atom, or, in other words, to their correlation, is of great importance. To take into account these correlations, some special methods and approaches were developed, such as the *R*-matrix [3], random phase approximation with exchange (RPAE) [4], many-body perturbation theory [5], etc. It was demonstrated that strong collective effects exist in atoms [4]. To take them into account, extensive calculations were performed, which required special computing codes [6], and reasonably good results were obtained. But owing to the complicated nature of the calculations, these results are far from being qualitatively transparent and, therefore, not physically clear.

Another approach to atomic structure is to treat it as a gas of electrons, confined in some volume in the Coulombic field of the nucleus. It is assumed that the properties and characteristics of the electronic shell are such that it behaves like a confined homogeneous electron gas of the same density. This approach, called Thomas-Fermi model [7], permits calculating the electron density as a function of the distance from the nucleus, and calculating the binding energy and other characteristics. As a result of these efforts, a method was developed which describes the atoms more or less adequately, but on the average, i.e. it completely neglects the specificity of each atom and its shell structure. The advantages of this approach are its lucidity and simplicity.

The aim of this paper is to simplify the picture further by applying some of the ideas of Thomas-Fermi model to a number of problems of the hydrogen ion in a strong electric field. The main aim here is not the accurate calculation of ground-state properties such as energy. Instead, we shall concentrate on problems whose very formulation is extremely difficult in the framework of more precise approaches (e.g., Hartree-Fock approach). Indeed, it will be demonstrated that the simple analytical method proposed can be quite naturally used in describing the possible collective motion of the electrons in an ion of hydrogen as well as the origin of a barrier, which makes he hydrogen ion stable, and its dependence on the amplitude and the frequency of the electric field.

## 2. Hydrogen atom

Let us start with a simple problem of a free hydrogen atom [8]. More complicated problems of a negative hydrogen ion in an external electric field we will consider later.

Wave function of electron in the ground state of a hydrogen atom is [9]

$$\psi(r) = [\exp(-r/2R)]/2(2\pi)^{1/2}R^{3/2}, \tag{1}$$

where *r* is the radius-vector and *R* is related to the equilibrium or not size of the electronic shell.

The energy of the ground state is [9]

$$W_0 = -(me^4/2\hbar^2), \tag{2}$$

where *e* is the electron (negative) charge, *m* is the electron mass and $\hbar$ is the Planck constant divided by $2\pi$.

The electrostatic field created by the electron shell, which electric charge density is

$$\rho(r) = e\psi^*(r)\psi(r) = (e/8\pi R^3)\exp(-r/R), \tag{3}$$

could be calculated using Gauss theorem. The corresponding electrostatic potential is

$$\varphi(r) = (e/r) - (e/2R)[(2R/r) + 1]\exp(-r/R). \quad (4)$$

In the vicinity of $r = 0$ it has the following form:

$$\varphi(r) \approx (e/2R) - (e/12R^3)r^2 + \ldots . \quad (5)$$

Kinetic energy of the electron is a quantum mechanically averaged value of its operator,

$$T = -(\hbar^2/2m)\Delta. \quad (6)$$

It is described by the following relation:

$$\langle T \rangle = \hbar^2/8mR^2. \quad (7)$$

So the total energy of the electronic shell and the positively charged core placed in the center of the electronic shell is as follows:

$$W(R) = (\hbar^2/8mR^2) - (e^2/2R). \quad (8)$$

The electrostatic energy of the electronic shell itself should not be taken into account as it represents the so-called self-action. Energy, described by Eq. (8) has a minimum at $R = R_e = \hbar^2/2me^2$, as is well known [9]. The minimum value of the energy is given by Eq. (2).

In order to see the nature of the restoring forces, arising after the shift of the electronic shell, let us expand the energy $W(R)$ in a series around $R = R_e$ and $s = 0$ ($s$ is the shift of the electronic shell as a whole relative to the positively charged core). Using Eqs. (5,8), one can see that this expansion has the following form:

$$W(x, R - R_e) = -(me^4/2\hbar^2) + (e^2/2\alpha)s^2 + (2m^3e^8/\hbar^6)(R - R_e)^2 + \ldots , \quad (9)$$

where $\alpha$ is the polarizability of the hydrogen atom.

In classical mechanics the model of the movement of the electronic shell as a whole relative to the charged core does not yield the correct value of the polarizability [10,11]. It yields $\alpha = 6R_e^3 = 3\hbar^6/4m^3e^6$ for the hydrogen atom [8]. This value represents that part of the polarizability, which is coming from the bulk of the electronic shell (from the core of it) [9]. As to the experimentally observed polarizability, the tail of the electronic shell gives the main contribution to it, which is not described correctly by this classical model. The correct quantum mechanical value is $\alpha = 9\hbar^6/2m^3e^6$ [9].

The first term in the left-hand part of Eq. (9) is the energy of the ground state. The second term describes the potential of the harmonic oscillator (oscillatory dipole movement of the electronic shell as a whole around the positively charged core) in terms of the classical mechanics. The third term describes in the same framework the so-called breathing mode, that is the isotropic oscillations of the radius of the electronic shell $R$ around its equilibrium value, or the oscillations of the atomic size [10]. Being of a spherical symmetry, such motion is not accompanied by creation of a time-dependent dipole, quadrupole or any other multipole moments and therefore cannot emit electromagnetic radiation. In quantum mechanics the breathing mode for the hydrogen atom does not exist. In many electron atoms it could be realized as a collective motion of many electron system [10]. In quantum mechanics the only energy levels of the electron in a hydrogen atom are

$$W_\nu = -(me^4/2\nu^2\hbar^2), \quad \nu = 1, 2, 3, \ldots. \tag{10}$$

Now we shall start discussing more complicated problems.

## 3. Electrostatic energy of a negative hydrogen ion in an external field

Let us assume that the charge of the two electrons is confined in some volume restricted by the surface of a sphere and, for simplicity, is distributed homogeneously throughout this volume, which we shall call the volume of the ion. The concept of the size of the atom (ion) has long ago being considered applicable in atomic physics [12]. The ionic binding energy, $W_e$, is given as the sum of the kinetic energy, $T$, and the potential energy, which in fact is the electrostatic energy, $U$, of the ion. In this section we shall calculate $U$ for the negative hydrogen ion. To do this let us consider first two electrons confined in a sphere of a radius $R$. The electrostatic potential of these electrons is represented by the following equation [13]:

$$\varphi(r) = (3e/R) - (e/R^3)r^2, \tag{11}$$

where $r$ is the distance from the center of the sphere.

Now let us put the nucleus (proton) in the center of the sphere. Since the charge of the nucleus is $-e$, the interaction energy of the nucleus with the electronic shell is $-e\varphi(0) = -(3e^2/R)$. The electrostatic potential of all the particles inside the ion is

$$\varphi_{ep}(r) = (3e/R) - (e/R^3)r^2 - (e/r). \tag{12}$$

The electrostatic energy of the ion consists of the electrostatic energy of the electronic shell, the homogeneously charged ball of a radius $R$, electric charge $q = 2e$, and the interaction energy of the nucleus, point charge $-e$, with the electronic shell. The electrostatic energy of the charged ball is $(3/5)q^2/R = (12/5)e^2/R$ [13]. It consists of the interaction energy of two electrons and the electrostatic energy of two electrons themselves, $2\times 0.6(e^2/R) = 1.2(e^2/R)$. This last term represents self-interaction and should be subtracted from the electrostatic energy of the ball, $2.4e^2/R$. The interaction energy (see Eq. (11)) is $-3e^2/R$. So the total electrostatic energy of the ion in zero electrostatic field is $U(R) = -1.8e^2/R$ [11].

As was mentioned above, the self-interaction energies of the electrons and the nucleus should not be included in the total electrostatic energy of the ion.

It should be mentioned that both electrons in the ground state of the ion occupy the same ground level, their spins being antiparallel (the total spin is zero). For such an object one could expect that the accepted model would yield quite reasonable results.

Application of the external electric field, **E**, leads to the displacement of the electronic shell, $s$, relative to the nucleus in the direction opposite to that of the field. We assume that the electronic shell preserves its spherical shape and the uniform distribution of the electric charge. According to Eq. (11) regarded shift causes some change in the interaction energy of the electronic shell and the nucleus, $e^2s^2/R^3$. The change in the energy of the electrons due to the shift in a homogeneous external applied field, $-2esE$, also should be taken into account. The value of both terms mentioned, $(e^2s^2/R^3) - 2esE$, reaches the minimum value, $-R^3E^2$, at the state of equilibrium, when $s_e = (R^3/e)E$. The total electrostatic energy of the ion includes also that in zero field, $-1.8(e^2/R)$, calculated before. So finally we have [11]

$$U(R) = -1.8(e^2/R) - R^3E^2. \tag{13}$$

It should be mentioned here that the polarizability of the ion is $\alpha = p/E = R^3$ (where $p = es_e$ is the value of the dipole moment of the ion). The equilibrium value of the radius of a sphere, $R_e$, will be calculated in the next section.

## 4. Equilibrium ionic size

The competition between the electrostatic and kinetic energies determines the equilibrium ionic size. In our model the ion is a sphere with a specific radius, $R$, and the charge of the electrons is distributed homogeneously in this sphere. The kinetic energy, $T$, of a gas, confined in a volume $(4/3)\pi R^3$, and calculated in accordance with the ideas of Thomas-Fermi model, is [10]

$$T(R) = 1.10 N^{5/3} \hbar^2 / mR^2, \tag{14}$$

where $m$ is the electron mass and the number of the electrons in the ion, $N = 2$ ($1.10 N^{5/3} = 3.51$).

To describe experimental data properly, correction factors were introduced into Thomas-Fermi approximation a long time ago [14]. Let us therefore introduce a correction factor, $g$, into the kinetic energy of the ion:

$$T(R) = 3.51 g \hbar^2 / mR^2. \tag{15}$$

It is assumed that g < 1. The value of $g$ depends on the properties of the atom, which we wish to describe accurately. In this paper we shall use the correction factor to fit the total energy of the ion.

In Thomas-Fermi model the kinetic energy does not depend on the shape of a sample. For a bulk sample it depends on its volume only.

According to Eqs. (13,15), the total energy of the ion is

$$W(R) = 3.51(g\hbar^2/mR^2) - 1.8(e^2/R) - R^3E^2. \tag{16}$$

At $E = 0$ the equilibrium value of the radius of the ion, $R_{0e}$, corresponding to the minimum of $W(R)$ is $R_{0e} = 3.90 g\hbar^2/me^2$. This yields for the equilibrium value of the total energy, $W_{0e} = -0.231 me^4/g\hbar^2 = -(6.29/g)$ eV. Since the total energy of the hydrogen atom is $-13.598$ eV [15], and the electron affinity of a free electron to the hydrogen atom is 0.754 eV [15], the total energy of the ion is −14.35 eV. To obtain the correct value of the total energy we have to accept for the correction factor a value of $g = 0.438$. This value of the correction factor will be used later. At $E = 0$ we have $R_e = R_{0e} = 1.71 \hbar^2/me^2$. This value is 3.42 times larger than the radius of the hydrogen atom, $0.5\hbar^2/me^2$.

In the general case, the total energy, $W$, as a function of $R$, Eq. (16) and Fig. 1, decreases from infinity, and has a minimum at some $R = R_e$, after which it has a maximum at some $R = R_m > R_e$, and then drops to minus infinity.

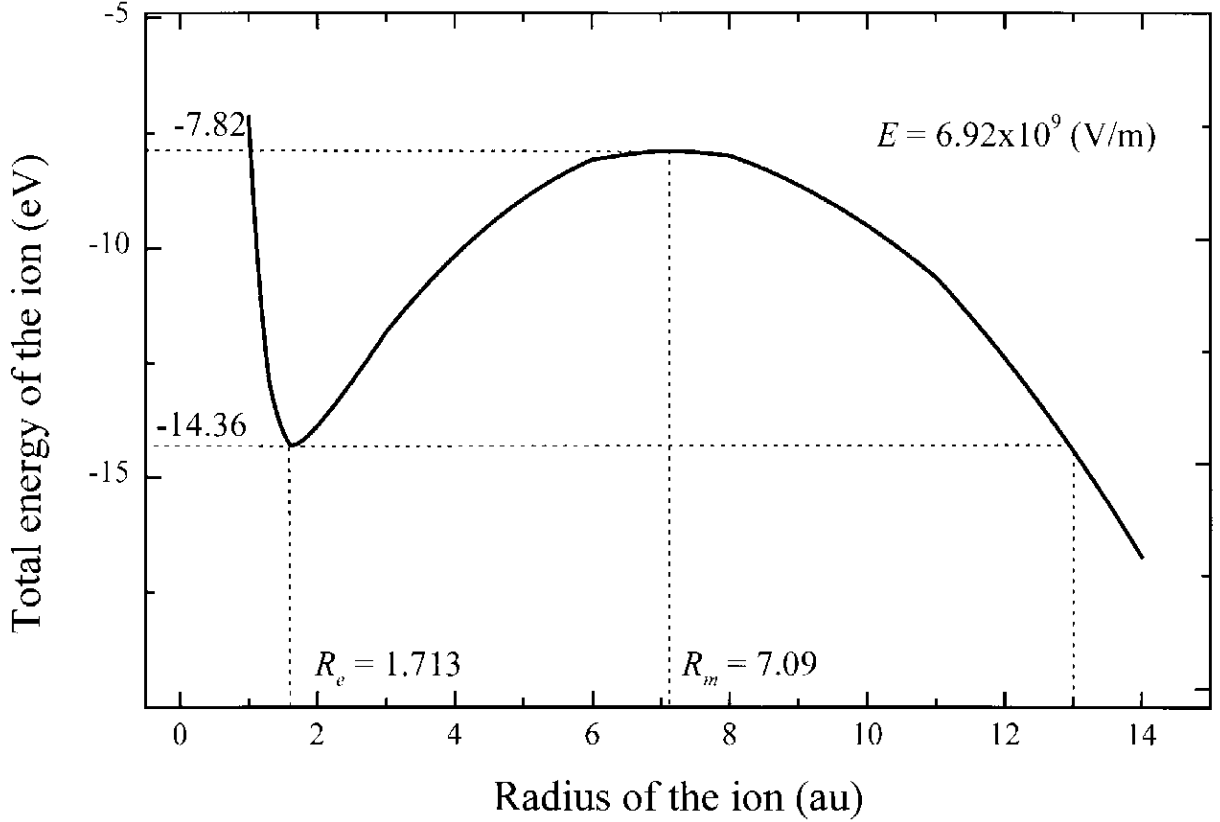

Figure 1. The total energy of the ion as a function of its radius for an external applied electric field E = 6.92×10$^9$ V/m.

The extremal values of $W$ are determined by the condition $\partial W(R)/\partial R = 0$, which yields the following equation for them:

$$(E/e)^2 R_{e,m}^5 - 0.6 R_{e,m} + 1.02(\hbar^2/me^2) = 0. \tag{17}$$

Between a minimum and a maximum, at $R = R_i$, the curve has a point of inflection, where the second derivative of the total energy is zero, $\partial^2 W(R)/\partial R^2 = 0$, which leads to the equation,

$$(E/e)^2 R_i^5 + 0.6 R_i - 1.54(\hbar^2/me^2) = 0. \tag{18}$$

With the increase in $E$ the height and width of the barrier decrease, and the size of the ion increases. It should be mentioned that the application of an external field leads to a slight decrease in the total energy of the ion. This decrease is a result of the interaction between the external field and the induced electrostatic moment.

At some large critical external field $E = E_c$, the maximum, the minimum and the point of inflexion coincide, and the ion loses its stability. Eqs. (17,18) yield the corresponding critical radius $R_c = 2.13\hbar^2/me^2$ and $E_c = -0.077m^2e^5/\hbar^4 = 4.38\times10^5$ V/m.

From Eq. (17) follows that

$$\partial R_e/\partial E^2 = (R_e^6/e^2)/[5.1246(\hbar^2/me^2) - 2.4R_e] > 0. \tag{19}$$

Since $2.4R_e < 5.1246\hbar^2/me^2$, this means that with the increase in $E$ from 0 to $E_c$ the value of $R_e$ increases concomitantly from $1.71\hbar^2/me^2$ to $2.13\hbar^2/me^2$ (almost 25% increase).

It is possible to consider another mechanism of the instability of the ion. The external field can be so strong that the equilibrium value of the displacement of the electronic shell relative to the nucleus, $s_e = (R^3/e)E$ (see **Section 3**), becomes larger than $R_e$ itself. This leads to a 'stripping' instability of the ion at a corresponding critical value of $E = E_s = e/R_e^2$. For the maximum value of $R_e$, $R_c = 2.13\hbar^2/me^2$, we have $E_s = 1.25\times10^6$ V/m. This is a rather high value of the critical electric field. We saw above that the expansion mechanism provides instability at a lower value of a critical field, $4.38\times10^5$ V/m. But we shall see below that the 'stripping' mechanism can be the actual one in a dynamic regime.

## 5. Dynamic model

To formulate dynamic model let us start with Newton equation for the electronic shell. External applied field acts on the electronic shell with the force $2eE$. As the shift of the electronic shell, $s$, causes some increase in the interaction energy of the electronic shell and the nucleus, $e^2s^2/R^3$, we have another force, acting on the electronic shell, $-2(e^2/R^3)s$. So Newton equation is as follows:

$$2m(\partial^2 s/\partial t^2) = 2eE - 2(e^2/R^3)s, \tag{20}$$

The radiation friction force [16] is not taken into account in Eq. (20). It can be neglected when the angular frequency, $\omega$, is considerably smaller than $3mc^3/4e^2 = 8\times10^{22}$ 1/s $= 5.27\times10^7$ eV (here $c$ is the speed of light in the vacuum). Neither shall we take into account the changes in the radius of the electronic shell, $R$, during oscillation, since we have seen above that the largest change possible is no more than 25%.

Let us assume that $E(t) = E_0\sin\omega t$. Then, as follows from Eq. (20), we have $s(t) = s_0\sin\omega t$ with

$$s_0 = eE_0/m(\omega_d^2 - \omega^2), \tag{21}$$

where $\omega_d(R) = (e^2/mR^3)^{1/2}$ is the frequency of the dipole oscillation of the electronic shell relative to the nucleus. For $R = R_0 = 1.71\hbar^2/me^2$ $\hbar\omega_d = 12.1$ eV. At the maximum displacement the kinetic energy of the oscillation is zero, the changes in the electrostatic energy of the ion in the external applied field is $e^2 s_0^2/R^3 - 2es_0 E_0$ (see **Section 3**), which can be written as

$$\delta U = e^2 E_0^2 (2\omega^2 - \omega_d^2)\omega_d^2/m(\omega_d^2 - \omega^2)^2. \tag{22}$$

When $\omega = 0$ Eq. (22) yields $\delta U = -R^3 E^2$ as expected. The maximum deviation reaches $R$ value at the 'stripping' threshold, when $E = E_s(\omega)$:

$$E_s(\omega) = mR(\omega_d^2 - \omega^2)/e. \tag{23}$$

When $\omega = 0$ Eq. (23) yields $E_c = e/R^2$ as expected (see **Section 4**). Eq. (23) shows that the higher the frequency the lower the 'stripping' threshold is.

## 6. Resonant instability

Now let us consider Newton equation, taking into account the radiation friction force, $(8e^2/3c^3)(\partial^3 s/\partial t^3)$ [16]:

$$2m(\partial^2 s/\partial t^2) = 2eE - 2(e^2/R^3)s + (8e^2/3c^3)(\partial^3 s/\partial t^3). \tag{24}$$

Let us assume that $E(t) = E_0 \sin\omega t$, and $s(t) = a\sin\omega t + b\cos\omega t$. Eq. (24) allows calculating the factors $a$ and $b$. The amplitude of the oscillation, $s_0 = (a^2 + b^2)^{1/2}$, is as follows

$$s_0 = -eE_0/m[(4e^2/3mc^3)^2 \omega^6 + (\omega_d^2 - \omega^2)^2]^{1/2}. \tag{25}$$

When $s_0 = R$ the 'stripping' instability occurs. This happens when $E_0 = E_s$ (the critical 'stripping' field),

$$E_s = -(mR/e)[(4e^2/3mc^3)^2 \omega^6 + (\omega_d^2 - \omega^2)^2]^{1/2}. \tag{26}$$

At the resonant frequency, $\omega = \omega_d$, we have $E_s = (4eR/3c^3)\omega_d^3$, which is only 0.443 V/m at $R = 1.71\hbar^2/me^2$ and $\hbar\omega_d = 12.1$ eV.

## 7. Discussion

A simple approach, based on Thomas-Fermi model was applied to describe some collective properties of the negative hydrogen ion. It was shown that for typical frequencies of solid-state laser ($\hbar\omega$ is about 1 eV) a static approach gives results close to the dynamic one. Corrections due to dynamic effects appeared to be rather small at these frequencies. Two mechanisms of the instability were considered. At low frequencies the expansion of the volume of an ion is the appropriate one. The corresponding critical field is about $4\times10^5$ V/m. The 'stripping' mechanism is very effective in the vicinity of the resonance. The critical 'stripping' amplitude of the electric field in the resonance is only about 0.443 V/m, but the frequency is rather high ($\hbar\omega_d$ is 12.1 eV).